\documentclass[unsortedaddress,prd,reprint,showpacs]{revtex4-1}

\usepackage{amsmath,empheq}
\usepackage{amsfonts}
\usepackage{amsthm}
\usepackage{amssymb}
\usepackage{graphicx}
\usepackage{hyperref}
\usepackage{soul}
\usepackage{url}
\usepackage{color}

	\newcommand{\ncd}{\newcommand}
	\ncd{\mrm}    {\mathrm}
	\ncd{\beq} {\begin{equation}}
	\ncd{\eeq} {\end{equation}}

	\def\d{{\rm d}}

\begin{document}
	\title{A thermostat algorithm generating target ensembles}
	\date{\today}

	\author{A. Bravetti}
        \email{bravetti@correo.nucleares.unam.mx}
	\affiliation{Instituto de Ciencias Nucleares, Universidad Nacional Aut\'onoma de M\'exico,\\ AP 70543, M\'exico, DF 04510, Mexico.}

	\author{D. Tapias}
	\email{diego.tapias@nucleares.unam.mx }
	\affiliation{Facultad de Ciencias, Universidad Nacional Aut\'onoma de M\'exico,\\ AP 70543, M\'exico, DF 04510, Mexico.}

	\begin{abstract}
	We present a deterministic algorithm called contact density dynamics that generates any prescribed target distribution in the physical phase space. Akin to the famous model of Nos\'e-Hoover, our algorithm is based on a non-Hamiltonian system in an extended phase space. However the equations of motion in our case follow from contact geometry and we show that in general they have a similar form to those of the so-called density dynamics algorithm. As a prototypical example, we apply our algorithm to 
produce Gibbs canonical distribution for a one-dimensional harmonic oscillator. 
	\end{abstract}


        \maketitle

        \section{Introduction}
Equilibrium statistical mechanics is a beautiful mathematical construction based on Gibbs canonical distribution and a very powerful tool that permits
to establish a link between the microscopic laws of motion and the macroscopically observable properties of systems with a large number of particles.
However, some conceptual and practical problems in this framework are still unsettled. 

A  major issue regards the mechanical foundations of the equilibrium distribution. In fact, the dynamical evolution of a Hamiltonian system is confined to a hypersurface 
of constant energy of the phase space and therefore the only possible distribution for the energy of the system from a dynamical perspective is a delta distribution, which represents the microcanonical ensemble.
Therefore a relevant problem at the foundations of statistical mechanics, which is also of primary practical importance for numerical simulations, is that of finding a well-defined dynamics that can lead
to ensembles which are different from the microcanonical one.
In this case several proposals have been found, which are generally based on defining a fictitious dynamical system in an extended phase space that reduces to the desired non-Hamiltonian dynamics
in the physical phase space, with the property that the invariant distribution reproduces a specified ensemble. Such algorithms are known in the literature as \emph{thermostat algorithms}.
The paradigmatic example is Nos\'e-Hoover algorithm (NH), which generates the canonical ensemble in the physical phase space~\cite{hoover1985canonical}  
 (see also~\cite{kusnezov1990canonical,Tuckerman2001,TuckermanBook,1998Morriss, evansbook, sergi2001non, sergi2003non} for further references).

In~\cite{fukuda2002tsallis} an algorithm based on NH idea that generates \emph{any} distribution on the physical phase space has been
proposed, which has been called \emph{density dynamics} (DD). Here we will introduce an algorithm similar in spirit to that of DD. The main difference is that
our procedure is motivated through a geometrical setting. 
In fact 
the systems that we are going to introduce are the natural extension of classical Hamiltonian systems 
to a space with an extra dimension and are known in the literature as \emph{contact Hamiltonian systems}~\cite{arnold2001dynamical,Boyer}. Their dynamics includes standard
Hamiltonian dynamics in some particular cases that we shall point out. 
However, in the general case it is more rich and we will show that this generality is the essential ingredient to allow for the dynamical generation of ensembles different from the microcanonical one.
For this reason we refer to our algorithm as \emph{contact density dynamics} (CDD).

To introduce our algorithm we proceed in three steps,  akin to the NH and DD procedures.
We start with a class of dynamical systems in an extended phase space, which in our case is given by {contact Hamiltonian systems}.
The second step is to find an invariant measure for their flow. 
This step has been pursued in~\cite{bravetti2015liouville}, where it was remarked the important fact
that there is a \emph{unique} invariant measure depending only on the generating function in the extended phase space. 
Finally, the last step is to show that by a proper choice of the generating function and by integrating out the additional unphysical degree of freedom, 
any desired distribution in the physical phase space can be generated.
We argue that, 
assuming that the dynamics in the extended phase space is ergodic, our results provide a dynamical foundation for different ensembles. 
To show that this is indeed the case, 
 we include a 
numerical simulation generating Gibbs canonical ensemble for a one-dimensional harmonic oscillator.

In what follows we will first introduce the basics of NH and DD algorithms and then present our proposal.
To fix the notation, we will always denote by $\Gamma$ the physical phase space, with variables $(p,q)$, where $p$ and $q$ are $n$-dimensional
vector and $n$ is the number of degrees of freedom of the system. Moreover, $\tilde\Gamma$ will indicate the extended phase space, a $(2n+1)$-dimensional
space with coordinates $(p,q,S)$.

  \section{Nos\'e-Hoover and density dynamics}
  \subsection{Nos\'e-Hoover}
The logic of the NH algorithm follows three steps.
   Step~1 is simply the definition of a dynamical system in $\tilde\Gamma$, given by
  \begin{empheq}{align}
	\label{NH1}
	& \dot {q}_{i} = \frac{\partial H(p,q)}{\partial p_{i}}\,,\\
	\label{NH2}
	& \dot{p}_{i} = -\frac{\partial H(p,q)}{\partial p_{i}}-S\,{p}_{i}  \,,\\
	\label{NH4}
	& \dot{{S}} =\frac{1}{Q}\left(\sum_{i=1}^{n}p_{i}\frac{\partial H(p,q)}{\partial p_{i}}-\frac{n}{\beta}\right)\,,
	\end{empheq}
   where 
   ${p}_{i}$ and ${q}_{i}$ are the physical positions and momenta,  
   $Q$ is a positive constant and $\beta=1/k_{\rm B}T$. Here $S$ is an additional variable introduced ad-hoc in order to generate a non-Hamiltonian dynamics on $\Gamma$ 
   with the desired property of having control over the temperature $T$. 
   Furthermore, $H(p,q)$ is the Hamiltonian of the system.
   Step~2 is the identification of an invariant measure on $\tilde\Gamma$. It turns out (see e.g.~\cite{Tuckerman2001}) that the system \eqref{NH1}-\eqref{NH4} has 
   the invariant measure

        \beq\label{muNHextended}
         \begin{split}
  \d\mu_{\rm NH}={\rm e}^{-\beta H({p},{q})}\,{\rm e}^{-\beta Q S^{2}/2}\, \d^{n}{p} \,\d^{n}{q}\, \d S\,,
	\end{split}
   \eeq
where  $\d^{n} p\, \d^{n}q \,\d S$ is the volume element of $\tilde\Gamma$.
   
 Step~3 consists in obtaining the corresponding measure on $\Gamma$ by integrating out the additional variable $S$.
 A direct integration in \eqref{muNHextended} gives (up to a multiplicative factor)
 \beq\label{muNHphysical}
         \begin{split}
 \left.\d\mu_{\rm NH}\right|_{\Gamma} =  {\rm e}^{-\beta H({p},{q})} \d^{n}{p} \,\d^{n}{q}\,,
	\end{split}
   \eeq
 which coincides with the canonical measure. This proves that NH dynamics can 
generate the canonical ensemble in the physical phase space, provided 
the dynamics \eqref{NH1}-\eqref{NH4}
is ergodic~\cite{kusnezov1990canonical,plastino1997dynamical}.

\subsection{Density Dynamics}
The DD algorithm aims to generalize the NH equations in order to yield any distribution on the physical phase space.
The key idea of DD is to define  an ad-hoc  dynamical system on $\tilde\Gamma$ with the property that its invariant distribution coincides with an arbitrary $\rho(p,q,S)$.  
Then $\rho(p,q,S)$ is projected to $\Gamma$ to
obtain the desired distribution.
To do so, one starts with the function 
	\beq\label{Theta}
	\Theta(p,q,S)=-{\rm ln} \rho(p,q,S)
	\eeq 
and writes the flow
	\begin{empheq}{align}
	\label{DD1}
	& \dot {q}_{i} =  \frac{\partial\Theta(p,q,S)}{\partial p_{i}}\,,\\
	\label{DD2}
	& \dot{p}_{i} =  -\frac{\partial\Theta(p,q,S)}{\partial q_{i}}-\frac{\partial\Theta(p,q,S)}{\partial S}\, {p}_{i} \,,\\
	\label{DD4}
	& \dot{{S}} =\sum_{i=1}^{n}p_{i} \frac{\partial\Theta(p,q,S)}{\partial p_{i}}-{n}\,.
	\end{empheq}
It can be checked then that Liouville equation ${\rm div} \rho X=0$ is satisfied, with $X$ the vector field generating the flow \eqref{DD1}-\eqref{DD4}.
Therefore	
$\rho(p,q,S)$ is the invariant distribution on $\tilde\Gamma$.
For instance, when $\Theta(p,q,S)=\beta[H(p,q)+Q S^{2}/2]$ one recovers the NH case with the distribution \eqref{muNHextended}.

A simple and very useful case is the one in which the invariant distribution $\rho(p,q,S)$ is of the form
	\beq\label{rhoproduct}
	\rho(p,q,S)=\rho_{\rm t}(p,q)\,f(S)\,,
	\eeq
where $\rho_{\rm t}(p,q)$ is the target distribution on $\Gamma$ and $f(S)$ is a normalized distribution for the thermostatting variable $S$.
Being \eqref{rhoproduct} a product of two independent distributions, the integration of the variable $S$ is straightforward and the result is the desired distribution $\rho_{\rm t}(p,q)$
in the physical phase space. 

In the following we will present our new algorithm for generating equilibrium ensembles. As for the above description of the NH algorithm, we will divide it into three steps
and we will show that, although it is derived from a geometric perspective, it retains all the positive features of the DD algorithm.

	\section{Contact density dynamics}
	
       \subsection{Step 1: Contact Hamiltonian systems}
       Contact Hamiltonian systems are defined in a precise geometric fashion starting from a generating function in the extended phase space  
       which we will indicate as $h(p,q,S)$.
       The function $h$ is called the \emph{contact Hamiltonian} of the system (for more details see e.g.~\cite{arnold2001dynamical,Boyer,bravetti2015liouville}).
       The properties of such systems have already been exploited in physics. 
       In particular, they are relevant in thermodynamics~\cite{Mrugaa:2000aa,eberard2007extension,TPSSASAKI,CONTACTHAMTD,2015JMP....56g3301G}
       and in control theory~\cite{favache2010entropy,ramirez2013feedback}.
       Recently, it has been also proposed that they can be suitable to study the statistical mechanics of nonconservative systems~\cite{bravetti2015liouville} 
       and to improve the efficiency of Monte Carlo simulations~\cite{2014arXiv1405.3489B}.
	For our discussion, it is sufficient to write down the dynamical equations thus generated, which read 
	\begin{empheq}{align}
	\label{z1}
	&  \dot{q}_{i} = \frac{\partial h(p,q,S) }{\partial p_i}\,,\\
	\label{z2}
	& \dot{p}_{i} =  -\frac{\partial h(p,q,S)}{\partial q_{i}} + \frac{\partial h(p,q,S)}{\partial S} \,p_i \,,\\
	\label{z3}
	& \dot S = - \sum_{i=1}^{n} p_i \frac{\partial h(p,q,S)}{\partial p_i}+h(p,q,S)\,.
	\end{empheq}
From equations \eqref{z1}-\eqref{z2} it is clear that this dynamics induces a standard Hamiltonian dynamics over the physical phase space
whenever the generating function $h$ does not depend on $S$.	
Besides, the similarity with the NH and DD equations is evident and will be made more concrete in the next 
section -- c.f.~equations \eqref{G1}-\eqref{G3}.

\subsection{Step 2: The invariant distribution for contact Hamiltonian systems}
Although the system \eqref{z1}-\eqref{z3}
is non-Hamiltonian and there is no conserved quantity in the general case,
it was found in~\cite{bravetti2015liouville} that
 there is only one invariant measure on $\tilde\Gamma$ which depends uniquely on $h$ whenever $h\neq 0$.
 This is given by
	\beq\label{CanInvariantMeasure}
	\d\mu =  \dfrac{|h|^{-(n+1)}}{\mathcal{Z}_{n}}  \, \d^{n} p\, \d^{n}q\,\d S\,,
	\eeq
where $\left|\cdot\right|$ is the absolute value and $\mathcal{Z}_{n}$ is the partition function.
Thus, equation  \eqref{CanInvariantMeasure} shows  that the invariant measure of the dynamics generated
 by any contact Hamiltonian system in the extended phase space has a power law distribution. 
We show below that for a proper choice of $h$, the invariant measure \eqref{CanInvariantMeasure} induces any desired distribution on $\Gamma$, just as in the DD case.

\subsection{Step 3: Integrating out S and recovering the target distribution}
Let us proceed as in the preceding discussion about DD and assume that we wish to induce the \emph{target} distribution $\rho_{\rm t}(p,q)$ on $\Gamma$.
Considering the measure \eqref{CanInvariantMeasure}, together with the choice of the contact Hamiltonian 
	\beq\label{generalh}
	h(p,q,S)=\left[{\rho_{\rm t}(p,q)}f(S)\right]^{-\frac{1}{n+1}}\,,
	\eeq
it turns out that the invariant distribution on $\tilde\Gamma$ is set to be \eqref{rhoproduct}.
Moreover, with the choice of $h$ as in \eqref{generalh}, the function $h$ is always positive and therefore the absolute value
in \eqref{CanInvariantMeasure} is not necessary and we avoid regions where $h=0$ and the invariant measure is degenerate. 

Now, since $f(S)$ is a normalized distribution by assumption, we can integrate out the unphysical degree of freedom $S$ and obtain the induced measure 
on $\Gamma$, which is
	\beq\label{inducedmeasure}
	\left.\d\mu\right|_{\Gamma} =  {\rho_{\rm t}(p,q)} \, \d^{n} p\, \d^{n}q\,.
	\eeq
This concludes our algorithm for generating any desired ensemble on the physical phase spce.

Notice that different choices of the target distribution lead to different $h$ in \eqref{generalh}
and therefore to different dynamical equations of the form \eqref{z1}-\eqref{z3}. Moreover,  \eqref{generalh} is not the only possibility for the generating function.
We decided to present this form for clearness because in this case it is particularly simple to integrate out $S$.
A comment on ergodicity is also at order. Since $h$ as in \eqref{generalh} is always greater than zero, the flow equations
\eqref{z1}-\eqref{z3} do not have any fixed points, which are obstructions to ergodicity.
Finally, from the form of $h$ as in \eqref{generalh}, the dynamical equations on $\tilde\Gamma$ take the form
	\begin{empheq}{align}
	\label{G1}
	& \dot {q}_{i} =  \frac{h}{n+1}\frac{\partial\Theta(p,q,S)}{\partial p_{i}}\,,\\
	\label{G2}
	& \dot{p}_{i} =  \frac{h}{n+1}\left[-\frac{\partial\Theta(p,q,S)}{\partial q_{i}}+\frac{\partial\Theta(p,q,S)}{\partial S}\, {p}_{i}\right] \,,\\
	\label{G3}
	& \dot{{S}} =\frac{h}{n+1}\left[-\sum_{i=1}^{n}p_{i} \frac{\partial\Theta(p,q,S)}{\partial p_{i}}+{n}+1\right]\,,
	\end{empheq}
where $\Theta$ is given by \eqref{Theta}.
These equations suggest that the CDD algorithm is a re-scaling of the DD algorithm on the extended phase space by the positive function $h/(n+1)$. 
The relationship between CDD and DD is beyond of the scope of this work and it will be explored in future efforts.
Having established our algorithm, in the next section we will apply it to a concrete example, the generation of 
Gibbs canonical distributions for a one-dimensional harmonic oscillator.

%


 \section{Numerical simulation}
 In this section we consider a one-dimensional harmonic oscillator and show that our algorithm produces Gibbs canonical distribution in the physical phase space.
 This is a standard test for thermostat algorithms~\cite{tuckerman1992chains, kusnezov1990canonical,Tuckerman2001,TuckermanBook,1998Morriss}. For instance,
 it has been shown that the NH equations cannot generate Gibbs ensemble for this system
 due do the lack of ergodicity~\cite{hoover1985canonical,kusnezov1990canonical, legoll2007nonergodicity,watanabe2007ergodicity}.

Following equation \eqref{generalh}, the contact Hamiltonian for this system is
	\beq
	h(p,q,S) = \left(\dfrac{{\rm e}^{-\beta H(p,q)}}{\mathcal{Z}} f(S)\right)^{-1/2} ,
	\eeq
with $H(p,q)$ the Hamiltonian function of a harmonic oscillator with potential $U(q) = 2q^2$,
$\mathcal Z=\pi/\beta$ the corresponding partition function 
and $f(S)$ a normalized distribution. 
The freedom in  $f(S)$ allows us to do numerical tests for different distributions and choose the most
 adequate according to the ergodicity of the corresponding dynamical system and to the computational cost of the numerical integration of the equations of motion. 
Considering these issues, we select $f(S)$ to be the logistic distribution with scale $1$ and mean $c$, that is 
	\beq
	f(S;c) = \frac{{\rm e}^{S-c}}{(1+{\rm e}^{S-c})^2} \, . 
	\eeq
The choice of the numerical value of $c$ is also guided by the same principles mentioned above (ergodicity and computational cost). 
For the simulation we fix $c=2$, $k_B = m = 1$  and $\beta = 0.1$.

To integrate the equations of motion we use the Taylor series method for ODEs~\cite{corliss1982taylor,barrio2005taylor, jorba2005taylor}
and we implement it by means of a Julia code made available at~\cite{juliacode}.
The order in the Taylor series is equal to 28, thus the local error at each step is of the order of $O((\Delta t)^{29})$. 
The method uses a variable stepsize and a tolerance of $1.0 \times 10^{-20}$. 
This procedure is particularly efficient for high-accurate computations in low dimensional systems and therefore it is appropriate for our problem~\cite{barrio2005taylor}.
Since the stepsize in the method is not constant, 
we need to fix a sample time. We choose 
	$\Delta t_{\rm sample}=0.05$
and we decide to stop the simulation after a number of samplings $n_{\rm sampling}=1\times 10^{6}$, which corresponds to a total integration time 
$t_{\rm total} = \Delta t_{\rm sample}\times n_{\rm sampling}=5\times 10^{4}$.

\begin{figure}[h!]
\includegraphics[scale=0.5]{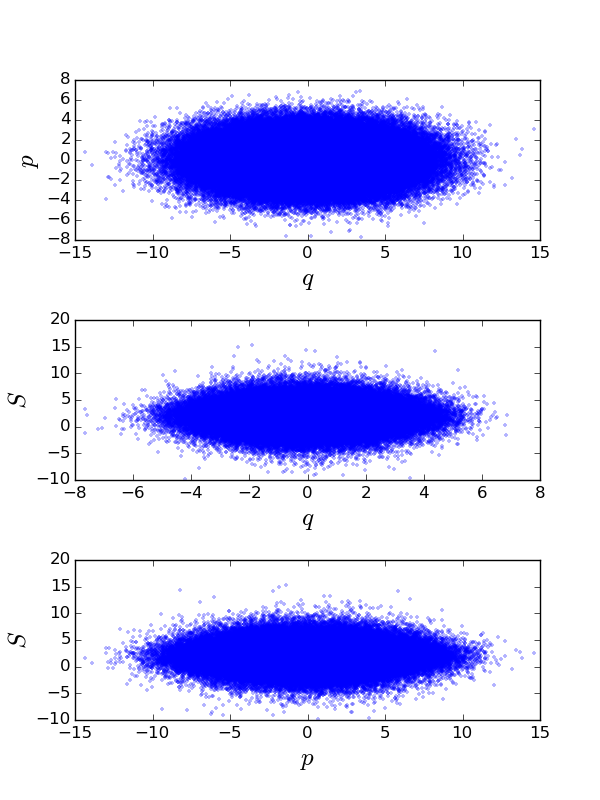}
\caption{Projections of the extended phase space orbit of a one-dimensional harmonic oscillator in the $(p,q)$, $(S,q)$ and $(S,p)$ planes. 
{An orbit with $3 \times 10^5$ points is shown. Initial condition $[q_0, p_0, S_0]$ $=$ $[0.12578471404894542,0.7479637648489665,$ $0.917435858684718]$. More details in the text.}}
\label{projections}
\end{figure}

In figure~\ref{projections} we display the projections to different planes of the orbit of the system with a 
 randomly generated 
initial condition. 
We see that the phase space is filled by the orbit.
We have analyzed the orbits of $10^3$ different random initial conditions and checked that the filling of the phase space is a generic property, which suggests the ergodicity of the system. 

\begin{figure}[h!]
\includegraphics[scale=0.45]{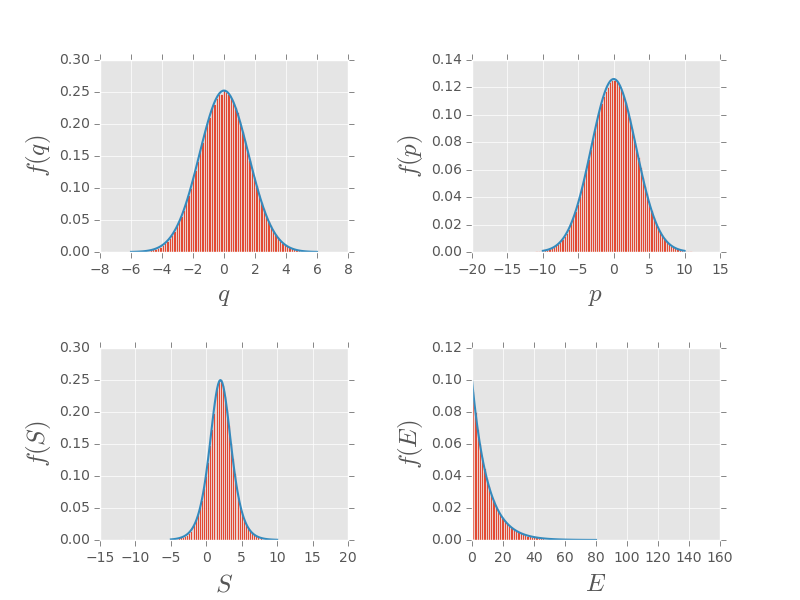}
\caption{Histograms of the frequencies for $q,p,S$ and $E$ and corresponding theoretical distributions (solid line) 
for a one-dimensional harmonic oscillator. The initial condition is the same as for figure \ref{projections}. More details in the text.}
\label{histograms}
\end{figure}

\begin{figure}[h!]
\includegraphics[scale=0.45]{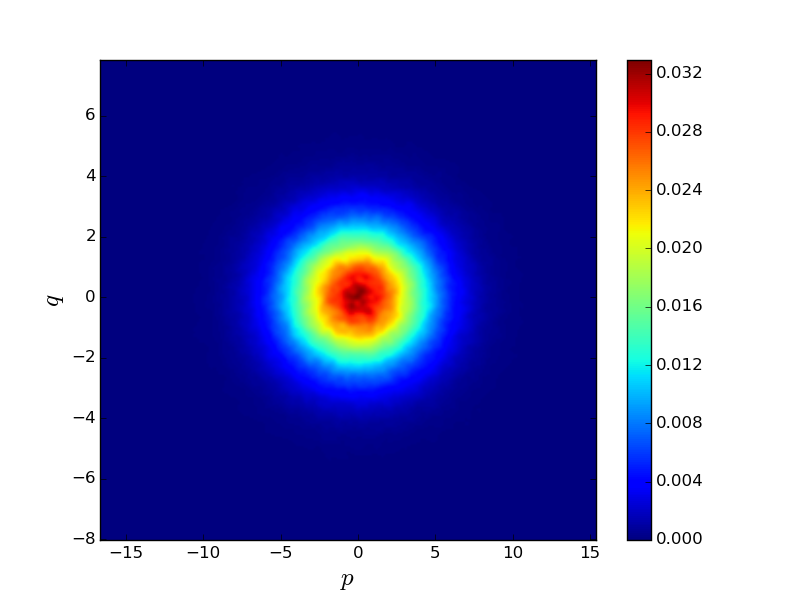}
\caption{Numerical joint probability distribution for $p$ and $q$ for a one-dimensional harmonic oscillator. The initial condition is the same as for figure \ref{projections}.
More details in the text.}
\label{joint}
\end{figure}

In figure~\ref{histograms} we show the histograms of the frequencies of the numerical values of $q,p,S$ and $E = H(p,q)$ for the specified 
trajectory and compare them with their theoretical distributions. 
The histograms are in good agreement with the theoretical curves.
Figure~\ref{joint} displays the numerical joint probability distribution of $p$ and $q$ along the orbit. 
The Gaussian character of the bivariate distribution is clearly observed.
All these tests indicate that the CDD correctly generates Gibbs distribution for this system.

\begin{figure}[h!]
\includegraphics[scale=0.45]{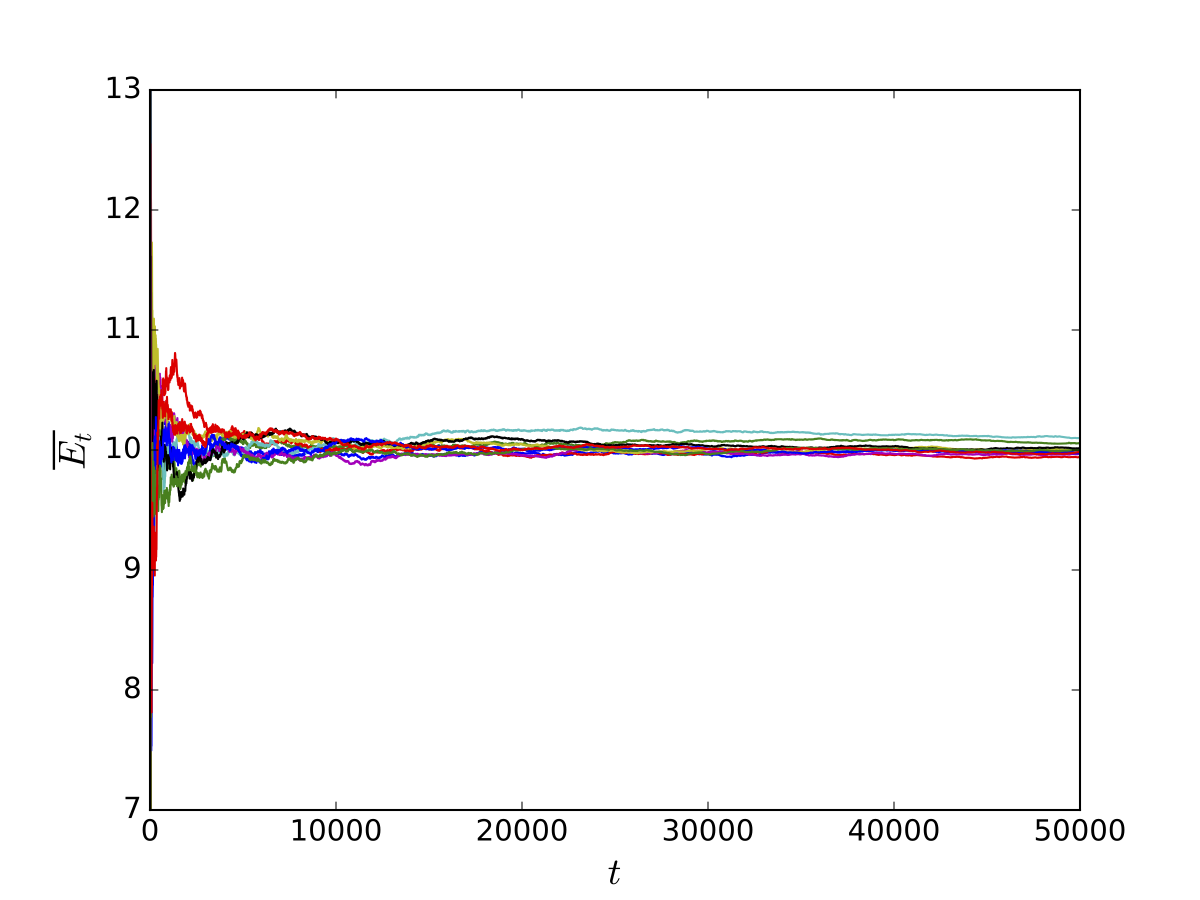}
\caption{Time average of the energy as a function of time for $10$ different initial conditions. Details in the text.}
\label{timeavg}
\end{figure}

As a final examination, we compute the time averages of the energy $\overline{E_t}$ for an \textit{ensemble} of $10^2$ oscillators and compare them with the ensemble average 
$\langle{E}\rangle = {1}/{\beta} = 10.0$. At the final time $t_{\rm total}$ the relative error for each element of the ensemble is less than $2 \%$. 
In figure~\ref{timeavg} we plot the evolution of $\overline{E_t}$ for 
$10$ representative elements.
The convergence of the time averages to the ensemble average of the energy is a further indication of the ergodicity of the system.

\section{Conclusions}
Hamiltonian mechanics and symplectic geometry are at the foundations of equilibrium statistical mechanics of conservative systems since they produce the microcanonical ensemble.
Here we have proposed an algorithm based on contact geometry and the corresponding Hamiltonian systems that dynamically produces any desired ensemble.
This might provide a theoretical basis for the equilibrium statistical mechanics of nonconservative systems.

We have shown that our algorithm generates equations of motion which have the same structure of those provided by density dynamics. 
However,
the main difference between our algorithm and DD is that our framework is grounded on
 the geometry of the extended phase space.

To investigate the ergodicity of the dynamics induced by our algorithm and prove that it effectively yields the desired target distribution in the physical phase space,
we have presented an example in which we simulated a one-dimensional harmonic oscillator in Gibbs canonical ensemble.
We have considered different curves in the phase space, marginal and joint distributions and the time averages of the energy for several randomly generated initial
conditions. From all these tests we
conclude that the system is in the canonical ensemble, as expected. 


In future works we wish to clarify in detail the relationship between our algorithm and density dynamics and to study the physical relevance of the contact
Hamiltonian. 
Moreover, we 
are going to apply the present proposal to construct several systems in
different ensembles.
We consider that  our contact density dynamics algorithm can be useful in the design of molecular dynamics simulations
 and that it establishes a step forward in the theoretical understanding of equilibrium in nonconservative systems.

\section*{Acknowledgements}
The authors would like to thank C. S. L\'opez-Monsalvo, C. Gruber, M. Seri, L. Benet and D. Sanders for insightful discussions.
AB acknowledges financial support from the A. della Riccia Foundation (postdoctoral fellowship). DT acknowledges financial support from CONACyT, CVU No. 442828.

\bibliographystyle{unsrt}
\bibliography{GTD.bib}

\begin{thebibliography}{10}

\bibitem{hoover1985canonical}
William~G Hoover.
\newblock Canonical dynamics: equilibrium phase-space distributions.
\newblock {\em Physical Review A}, 31(3):1695, 1985.

\bibitem{kusnezov1990canonical}
Dimitri Kusnezov, Aurel Bulgac, and Wolfgang Bauer.
\newblock Canonical ensembles from chaos.
\newblock {\em Annals of Physics}, 204(1):155--185, 1990.

\bibitem{Tuckerman2001}
M.~E. Tuckerman, Y.~Liu, G.~Ciccotti, and G.~J. Martyna.
\newblock Non-{H}amiltonian molecular dynamics: Generalizing {H}amiltonian
  phase space principles to non-{H}amiltonian systems.
\newblock {\em The Journal of Chemical Physics}, 115(4):1678--1702, 2001.

\bibitem{TuckermanBook}
M.~E. Tuckerman.
\newblock {\em Statistical {M}echanics: {T}heory and {M}olecular {S}imulation}.
\newblock Oxford graduate texts. Oxford University Press, Oxford, 2010.

\bibitem{1998Morriss}
G.~P. Morriss and C.~P. Dettmann.
\newblock Thermostats: Analysis and application.
\newblock {\em Chaos: An Interdisciplinary Journal of Nonlinear Science},
  8(2):321--336, 1998.

\bibitem{evansbook}
D.~Evans and G.~Morriss.
\newblock {\em Statistical {M}echanics of {N}onequilirium {L}iquids}.
\newblock Cambridge {U}niversity {P}ress, 2008.

\bibitem{sergi2001non}
Alessandro Sergi and Mauro Ferrario.
\newblock Non-{H}amiltonian equations of motion with a conserved energy.
\newblock {\em Physical Review E}, 64(5):056125, 2001.

\bibitem{sergi2003non}
Alessandro Sergi.
\newblock Non-{H}amiltonian equilibrium statistical mechanics.
\newblock {\em Physical Review E}, 67(2):021101, 2003.

\bibitem{fukuda2002tsallis}
Ikuo Fukuda and Haruki Nakamura.
\newblock Tsallis dynamics using the {N}os{\'e}-{H}oover approach.
\newblock {\em Physical Review E}, 65(2):026105, 2002.

\bibitem{arnold2001dynamical}
V.~I. Arnold and S.~P. Novikov.
\newblock {\em Dynamical Systems IV: Symplectic Geometry and Its Applications}.
\newblock Dinamicheskie sistemy. Springer, 2001.

\bibitem{Boyer}
C.~P. {Boyer}.
\newblock {Completely integrable contact Hamiltonian systems and toric contact
  structures on $S^{2}\times S^{3}$.}
\newblock {\em {SIGMA, Symmetry Integrability Geom. Methods Appl.}}, 7:058, 22,
  2011.

\bibitem{bravetti2015liouville}
A.~Bravetti and D.~Tapias.
\newblock Liouville{'}s theorem and the canonical measure for nonconservative
  systems from contact geometry.
\newblock {\em Journal of Physics A: Mathematical and Theoretical},
  48(24):245001, 2015.

\bibitem{plastino1997dynamical}
AR~Plastino and C~Anteneodo.
\newblock A dynamical thermostatting approach to nonextensive canonical
  ensembles.
\newblock {\em Annals of physics}, 255(2):250--269, 1997.

\bibitem{Mrugaa:2000aa}
R.~Mruga{\l}a.
\newblock On a special family of thermodynamic processes and their invariants.
\newblock {\em Reports on Mathematical Physics}, 46(3), 2000.

\bibitem{eberard2007extension}
D~Eberard, BM~Maschke, and AJ~Van Der~Schaft.
\newblock An extension of {H}amiltonian systems to the thermodynamic phase
  space: Towards a geometry of nonreversible processes.
\newblock {\em Reports on mathematical physics}, 60(2):175--198, 2007.

\bibitem{TPSSASAKI}
A~Bravetti and CS~Lopez-Monsalvo.
\newblock Para-sasakian geometry in thermodynamic fluctuation theory.
\newblock {\em Journal of Physics A: Mathematical and Theoretical},
  48(12):125206, 2015.

\bibitem{CONTACTHAMTD}
A.~Bravetti, C.S. Lopez-Monsalvo, and F.~Nettel.
\newblock Contact symmetries and {H}amiltonian thermodynamics.
\newblock {\em Annals of Physics}, 361:377 -- 400, 2015.

\bibitem{2015JMP....56g3301G}
S.-I. {Goto}.
\newblock {Legendre submanifolds in contact manifolds as attractors and
  geometric nonequilibrium thermodynamics}.
\newblock {\em Journal of Mathematical Physics}, 56(7):073301, July 2015.

\bibitem{favache2010entropy}
Audrey Favache, Denis Dochain, and B~Maschke.
\newblock An entropy-based formulation of irreversible processes based on
  contact structures.
\newblock {\em Chemical Engineering Science}, 65(18):5204--5216, 2010.

\bibitem{ramirez2013feedback}
Hector Ramirez, Bernhard Maschke, and Daniel Sbarbaro.
\newblock Feedback equivalence of input--output contact systems.
\newblock {\em Systems \& Control Letters}, 62(6):475--481, 2013.

\bibitem{2014arXiv1405.3489B}
M.~J. {Betancourt}.
\newblock {Adiabatic Monte Carlo}.
\newblock {\em ArXiv e-prints}, May 2014.

\bibitem{tuckerman1992chains}
Glenn~J. Martyna, Michael~L. Klein, and Mark Tuckerman.
\newblock Nos\'e-{H}oover chains: {T}he canonical ensemble via continuous
  dynamics.
\newblock {\em The Journal of Chemical Physics}, 97(4), 1992.

\bibitem{legoll2007nonergodicity}
Fr\'ed\'eric Legoll, Mitchell Luskin, and Richard Moeckel.
\newblock Non-ergodicity of the {N}os\'e-{H}oover {T}hermostatted {H}armonic
  {O}scillator.
\newblock {\em Archive for {R}ational {M}echanics and {A}nalysis},
  184(3):449--463, 2007.

\bibitem{watanabe2007ergodicity}
Hiroshi Watanabe and Hiroto Kobayashi.
\newblock Ergodicity of a thermostat family of the nos{\'e}-hoover type.
\newblock {\em Physical Review E}, 75(4):040102, 2007.

\bibitem{corliss1982taylor}
George Corliss and Y.~F. Chang.
\newblock Solving {O}rdinary {D}ifferential {E}quations {U}sing {T}aylor
  {S}eries.
\newblock {\em ACM Trans. Math. Softw.}, 8(2):114--144, June 1982.

\bibitem{barrio2005taylor}
Roberto Barrio.
\newblock Performance of the {T}aylor series method for {O}{D}{E}s/{D}{A}{E}s.
\newblock {\em Applied Mathematics and Computation}, 163(2):525 -- 545, 2005.

\bibitem{jorba2005taylor}
\'Angel Jorba and Maorong Zou.
\newblock A {S}oftware {P}ackage for the {N}umerical {I}ntegration of
  {O}{D}{E}s by means of {H}igh-{O}rder {T}aylor {M}ethods.
\newblock {\em Experiment. Math.}, 14(1):99--117, 2005.

\bibitem{juliacode}
See supplemental material at
  \href{http://github.com/dapias/ContactFlowsTaylor}{http://github.com/dapias/ContactFlowsTaylor}
  for the {J}ulia code.

\end{thebibliography}

\end{document}